# Demonstration of long-lived high power optical waveguides in air

N. Jhajj, E. W. Rosenthal, R. Birnbaum, J.K. Wahlstrand, and H.M. Milchberg
*Institute for Research in Electronics and Applied Physics*
*University of Maryland, College Park, MD 20742*

**Abstract**

We demonstrate that femtosecond filaments can set up an extended and robust thermal waveguide structure in air with a lifetime of several milliseconds, making possible the very long range guiding and distant projection of high energy laser pulses and high average power beams. As a proof of principle, we demonstrate guiding of 110 mJ, 7 ns, 532 nm pulses with 90% throughput over ~15 Rayleigh lengths in a 70 cm long air waveguide generated by the long timescale thermal relaxation of an array of femtosecond filaments. The guided pulse was limited only by our available laser energy. In general, these waveguides should be robust against the effects of thermal blooming of extremely high average power laser beams.

## I. Introduction

Long range filamentation of intense femtosecond laser pulses in gases is an area of increasing interest, as it combines exciting potential applications with fundamental nonlinear optical physics [1, 2]. An intense pulse propagating in a transparent medium induces a positive nonlinear correction to the refractive index that co-propagates with the pulse as a self-lens. Once the laser pulse peak power exceeds a critical value, typically $P > P_{cr}$ ~5-10 GW in gases [1, 3], the self-induced lens overcomes diffraction and focuses the beam, leading to plasma generation and beam defocusing when the gas ionization intensity threshold is exceeded. The dynamic interplay between self-focusing and defocusing leads to self-sustained propagation of a tightly radially confined high intensity region accompanied by plasma of diameter <100μm [4] over distances greatly exceeding the optical Rayleigh range. Filaments can extend from millimeters to hundreds of meters, depending on the medium and laser parameters [1]. Among the applications of filaments are remote sensing [5], THz generation [6, 7], spectral broadening and compression of ultrashort laser pulses [8, 9], and channeling of electrical discharges [10]. Despite these applications, it remains a significant limitation that femtosecond filamentation cannot deliver high *average* power long distances in a single tight spatial mode. This is due to the fact that for laser pulses with $P$ ~ several $P_{cr}$, the beam will collapse into multiple filaments [11] with shot-to-shot variation in their transverse location. For $P_{cr}$ ~ 5-10 GW, this means that single filament formation requires pulses of order ~1 mJ. For a 1 kHz pulse repetition rate laser, this represents only 1 W of average power.

Here, we demonstrate a method employing filaments that can easily supersede this limitation by setting up a robust, long range optical guiding structure. It opens the possibility for optical guiding of hundreds of kilowatts of average power over long distances in the atmosphere. Aside from obvious directed energy applications [12], such guiding structures can greatly enhance applications such as atmospheric lasing [13], remote sensing [5, 14], and atmospheric laser communication [15].

## II. Results and discussion

### A. Gas hydrodynamics initiated by femtosecond filaments

Recently we found that the filament plasma, starting at a temperature and electron density of ~5eV and a few times $10^{16}$ cm$^{-3}$ [4], acts as a thermal source to generate long-lived gas density hole structures that can last milliseconds and dissipate by thermal diffusion [16]. Our earlier density measurements [16] were limited to ~40 µs resolution, so the early time dynamics on the nanosecond timescale had been simulated but not directly measured. These structures are initiated as the filament plasma recombines to a neutral gas on a ~10ns timescale. Owing to the finite thermal conductivity of the gas, the initial thermal energy invested in the filament plasma is still contained in a small radial zone, but it is repartitioned into the translational and rotational degrees of freedom of the neutral gas. The result is an extended and narrow high pressure region at temperatures up to a few hundred K above ambient. This pressure source launches a radial sound wave ~100 ns after the filament is formed. By ~1 µs, the gas reaches pressure equilibrium with an elevated temperature and reduced gas density in the volume originally occupied by the filament, after which the 'density hole' decays by thermal diffusion on a few millisecond timescale [16].

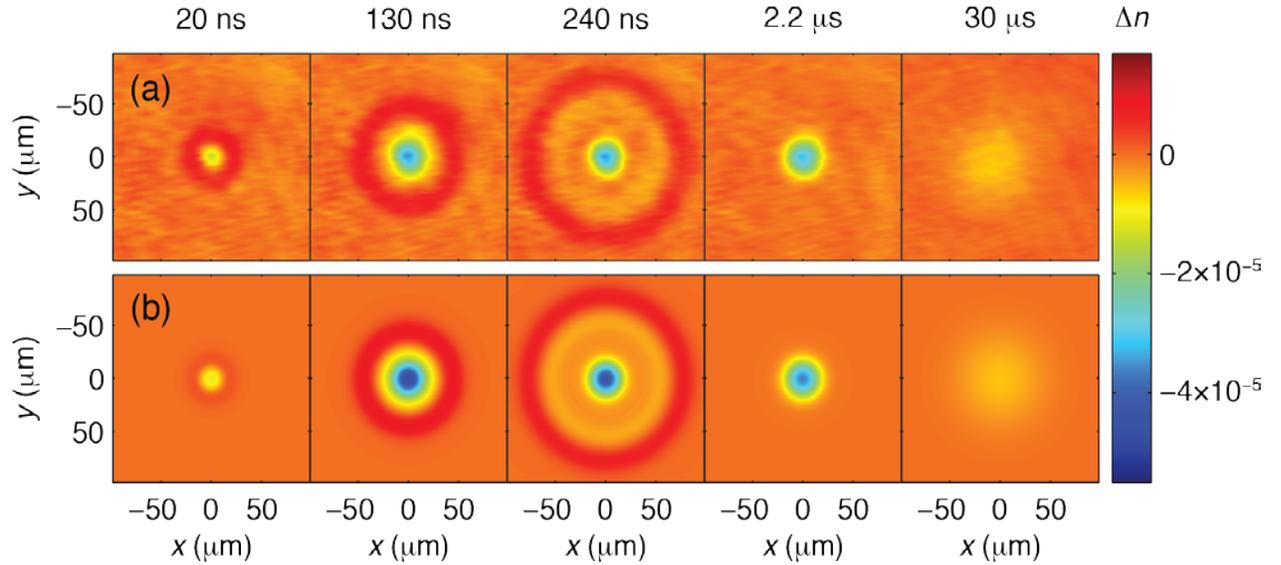

**Figure 1.** Gas dynamics following a single filament. (a) Interferometric measurement of the refractive index change following a short pulse as a function of the time delay of the probe pulse. (b) Hydrodynamic simulation, assuming initial filament electron temperature of 10 eV and electron density of $2\times10^{16}$ cm$^{-3}$.

The full dynamics are now clearly seen in Figure 1(a), which presents new, higher time resolution measurements of the 2D density hole evolution (expressed as air refractive index shift) of a short filament from nanoseconds through microseconds after filament formation. A 1D radial fluid code simulation, described in Appendix A, is shown in Fig. 1(b) for comparison and the results are in excellent agreement with the measurements. The experimental results verify that the density hole first deepens over tens of nanoseconds, and launches a sound wave which propagates beyond the ~200µm frame by ~300ns. By ~ 1-2 µs, pressure equilibrium is reached and the hole decays by thermal diffusion out to millisecond timescales.

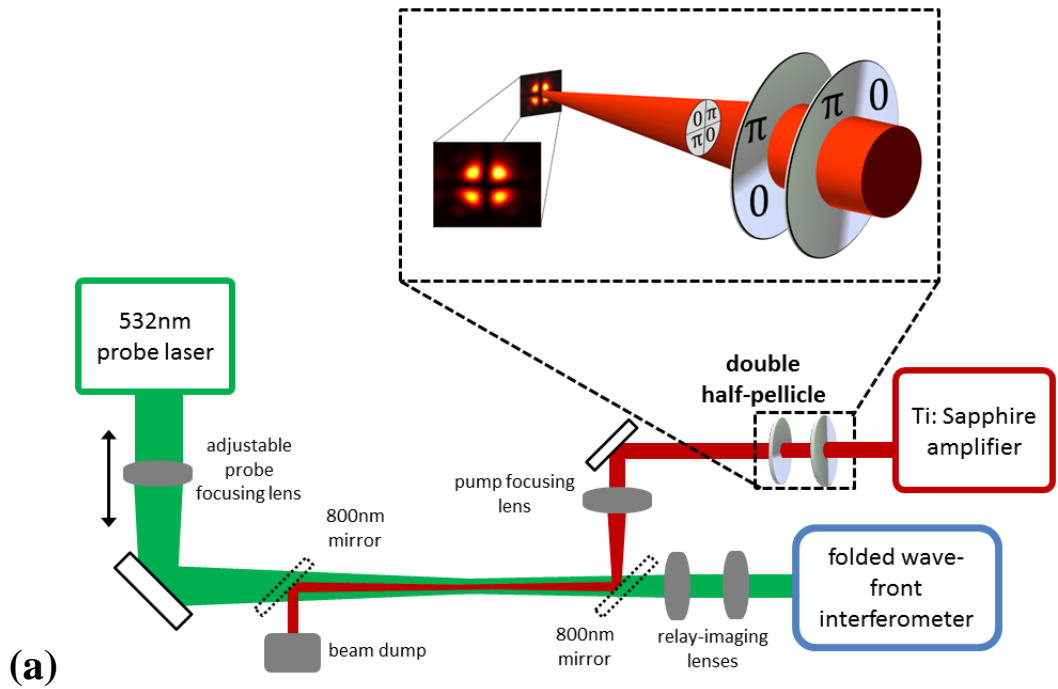

(a)

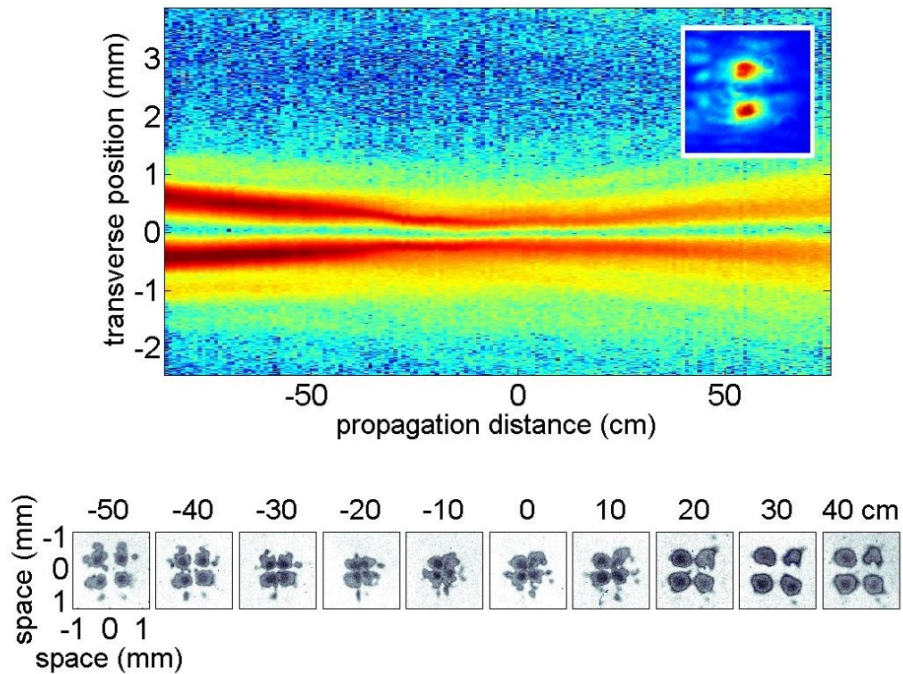

(b)

**Figure 2.** Generation of a filament array using half pellicles. (a) A 55 fs, 800 nm, 10 Hz pulsed laser is used to generate an array of four filaments. A pulse propagates through two orthogonal half-pellicles, inducing π phase shifts on neighbouring quadrants of the beam, and then are focused to produce a 4-filament with a TEM$_{11}$ mode (actual low intensity image shown). A 7 ns, 532 nm 10 Hz pulsed laser counter-propagates through the filament and is imaged either directly onto a CCD for guiding experiments or through a folded wavefront interferometer and onto a CCD for interferometry. (b) Rayleigh scattering as a function of $z$ with a bi-filament produced by a single half pellicle (the bi-filament far-field mode is shown in inset). The bottom row shows burn patterns produced by a 4-filament produced by two orthogonal half pellicles.

The gas density measurements were performed using the setup shown in Fig. 2(a) and described in Appendix B. A 532nm, 7ns laser pulse was used either as a longitudinal interferometric probe or as injection source for optical guiding. The transverse gas density profiles shown in Fig. 1(a), were obtained using the 532nm pulse as an interferometric probe of a single short ~ 2mm filament. The short filament length is essential for minimizing refractive distortion of the interferometric probe pulse [17].

We note that at no probe delay do we see an on-axis refractive index enhancement that might act as a waveguiding structure and explain a recent report of filament guiding [18]. At the longer delays of tens of microseconds and beyond, the thermal gas density hole acts as a negative lens, as seen in our earlier experiments [16].

**B. Multi-filament-induced guiding structure**

Although a single filament results in a beam-defocusing gas density hole, a question arises as to whether a guiding structure can be built using the judicious placement of more than one filament. We tested this idea with a 4-lobed focal beam structure using two orthogonal 'half-pellicles'. As seen in Fig. 2(a), the pellicles are oriented to phase-shift the laser electric field as shown in each near-field beam quadrant. Below the filamentation threshold, the resulting focused beam at its waist has a 4-lobed intensity profile as shown, corresponding to a Hermite-Gaussian TEM$_{11}$ mode, where the electric fields in adjacent lobes are $\pi$ phase shifted with respect to each other. Above the threshold, the lobes collapse into filaments whose optical cores still maintain this phase relationship and thus 4 parallel filaments are formed. As a demonstration of this, the top panel of Figure 2(b) shows an image of the Rayleigh side-scattering at 800 nm from a 2-lobed filament produced by a single half pellicle, indicating that the $\pi$ phase shift is preserved along the full length of the filament. This image was obtained by concatenating multiple images from a low noise CCD camera translated on a rail parallel to the filament. The images were taken through a 800 nm interference filter. The bottom panel shows burn patterns taken at multiple locations along the path of a ~ 70 cm long 4-lobe filament used later. For the 70 cm 4-filament, the filament core spacing is roughly constant at ~300μm over a $L$ ~ 50 cm region with divergence to ~ 1mm at the ends.

The effect of a 4-filament structure on the gas dynamics is shown in Fig. 3, a sequence of gas density profiles measured for a short ~ 2mm filament (produced at $f/35$) to minimize refractive distortion of the probe beam. The peak intensity was <$10^{14}$ W/cm$^2$, typical of the refraction-limited intensity in more extended filaments, so we expect these images to be descriptive of the gas dynamics inside much longer filaments. Inspection of the density profiles shows that there are two regimes in the gas dynamical evolution which are promising for supporting the guiding of a separate injected laser pulse. A shorter duration, more transient *acoustic regime* occurs when the sound waves originating from each of the four filaments superpose at the array's geometric centre, as seen in panel (a) of Fig. 3, causing a local density enhancement of approximately a factor of two bigger than the sound wave amplitude, peaking ~80 ns after filament initiation and lasting approximately ~50 ns. A far longer lasting and significantly more robust profile suitable for guiding is achieved tens of microseconds later, well after the sound waves have propagated far from the filaments. In this *thermal regime*, the gas is in pressure equilibrium [16]. As seen in panels (c) and (d) of Fig. 3, thermal diffusion has smoothed the profile in such a way that the gas at centre is surrounded by a 'moat' of lower density. The central density can be very slightly lower than the far background because its temperature is slightly elevated, yet it is still higher than the surrounding moat. The lifetime of this structure can be several milliseconds. In both the acoustic and thermal cases, the diameter of the air waveguide "core" is approximately half the filament lobe spacing. A movie of the 4-filament-induced gas evolution is provided in the supplementary material [19].

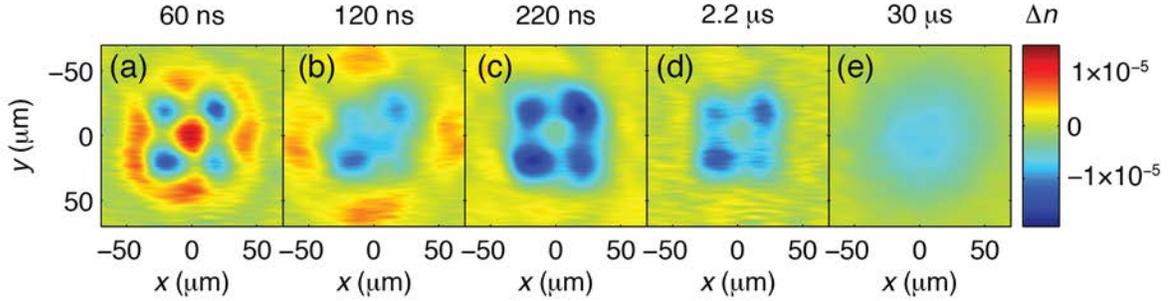

**Figure 3.** Interferometric measurement of the air density evolution induced by a 4-filament. (a) The acoustic waves generated by each filament cross in the middle, generating a positive index shift, producing the *acoustic guide.* (b) The acoustic waves propagate outward, leaving behind a density depression at the location of each filament. (c) The density depressions produce the *thermal guide,* with a higher central density surrounded by a moat of lower density. (d,e) The density depressions gradually fill in as the thermal energy dissipates. A movie of the 4-filament-induced gas evolution is provided in the supplementary material [19].

### C. Fibre analysis of air waveguides

Having identified two potential regimes for optical guiding, a short duration acoustic regime, and a much longer duration thermal regime, it is first worth assessing the coupling and guiding conditions for an injected pulse. We apply the fibre parameter $V$ for a step index guide [20] to the air waveguide, $V = (2\pi a/\lambda)\left(n_{co}^2 - n_{cl}^2\right)^{1/2} \sim (2\sqrt{2}\pi a/\lambda)(\delta n_{co} - \delta n_{cl})^{1/2}$, where the effective core and cladding regions have refractive indices $n_{co,cl} = n_0 + \delta n_{co,cl}$, $n_0$ is the unperturbed background air index ($n_0-1 = 2.77 \times 10^{-4}$ at room temperature and pressure [21]), $\delta n_{co}$ and $\delta n_{cl}$ are the (small) index shifts from background at the core and cladding, and the core diameter is $2a$, taken conservatively at the tightest spacing of the filament array. The numerical aperture of the guide is $NA = \lambda V/(2\pi a)$. Because accurate density profile measurements are restricted to short filaments, we use the results of Fig. 3 and apply them to much longer filaments that are inaccessible to longitudinal interferometry owing to probe refraction. As typical filament core intensities are restricted by refraction to levels $<10^{14}$ W/cm$^2$ [1], we expect that the measurements of Fig. 3 apply reasonably well to longer filaments and different lobe spacings. For the acoustic guide, we used a filamenting beam with lobe spacing of 150 μm, so $2a$ ~75 μm. Using $\delta n_{co}/(n_0-1) \sim 0.05$, and $\delta n_{cl}/(n_0-1) \sim -0.02$ from Fig. 3 then gives $V$~ 2.8 (> 2.405) and $NA$~ $6.3\times10^{-3}$, indicating a near-single mode guide with an optimum coupling f-number of $f/\#$ =0.5 /$NA$ ~ 80. For the thermal guide, we used a filamenting beam with lobe spacing of 300μm, so $2a$~150μm. From Fig. 3, the core index shift is $\delta n_{co}$ ~0 and the cladding shift is the index decrement at the moat, $\delta n_{cl}/(n_0-1) \sim -0.02$, giving $V$~2.9, corresponding to a near-single mode guide with $NA$~$3.2\times10^{-3}$, corresponding to $f/\#$ ~160.

**D. Injection and guiding experiments**

The experimental apparatus is depicted in Fig. 2(a). An end mode image from injection and guiding of a low energy $\lambda=532$ nm pulse in the acoustic waveguide produced from a 10 cm long 4-filament is shown in Fig. 4. In order to differentiate between guiding and the propagation of the unguided beam through the fully dissipated guide at later times (>2ms) we define the guiding efficiency as $(E_g - E_{ug})/(E_{tot} - E_{ug})$ where $E_g$ is the guided energy within the central mode, $E_{tot}$ is the total beam energy and $E_{ug}$ is the fraction of energy of the unguided mode occupying the same transverse area as the guided mode.

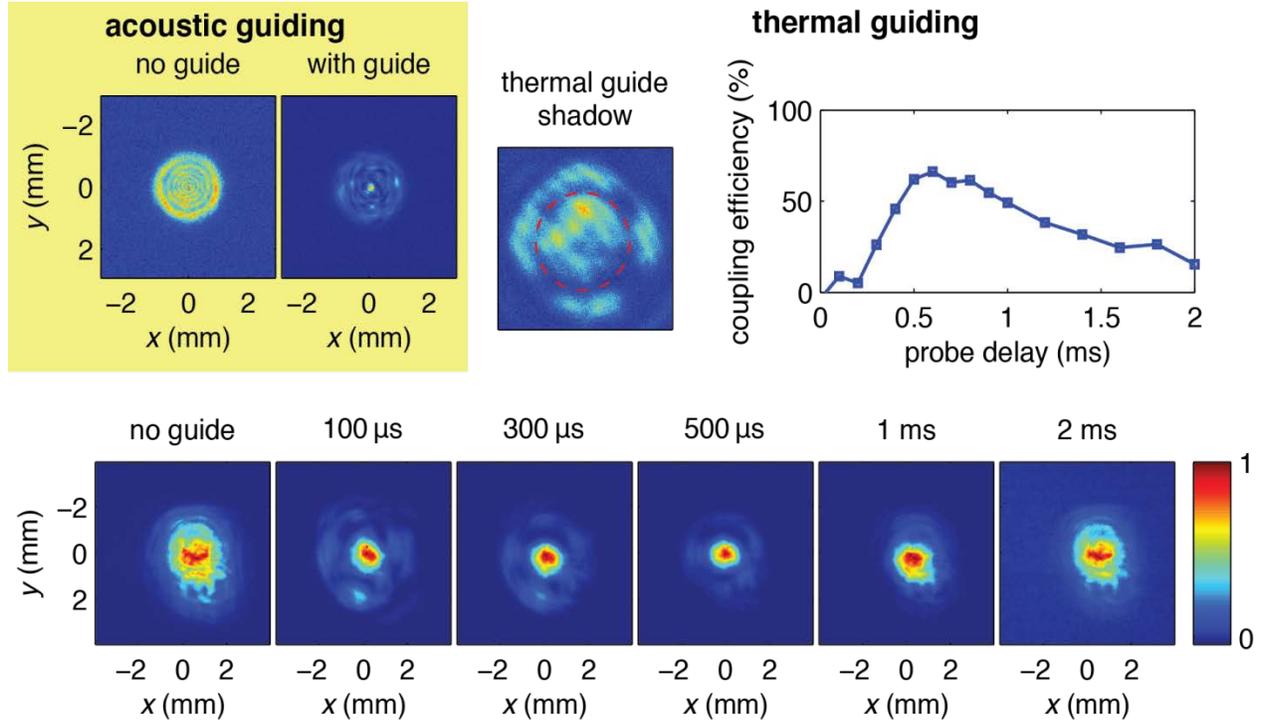

**Figure 4.** Demonstration of guiding of 7ns, $\lambda=532$ nm pulses in acoustic and thermal air waveguides produced by a 4-filament. The panel in the upper left shows the probe beam, which is imaged after the filamentation region, with and without the filament. The time delay of the probe was 200ns, which is in the acoustic guiding regime. The effect of the thermal waveguide, the shadow of which can be seen in the image in the top center (with a red dashed circle showing the position of the lower density moat), is shown in the bottom row, where the probe beam is imaged after the exit of the air waveguide with and without the filamenting beam. The coupling efficiency vs. injected pulse delay is shown in the upper right. Peak energy guided was ~110 mJ.

Best coupling occurred at an injection delay of ~200 ns and f/#>100, with a peak guided efficiency of 13%, although the guides were not stable on a shot-to-shot basis. Efficient guiding in the acoustic regime lasts only over an injection delay interval of ~100 ns, consistent with the time for a sound wave to cross the waveguide core region, $a/c_s$ ~ 100ns, where $2a=75\mu m$ and $c_s\sim3.4\times10^4$ cm/s is the air sound speed [22]. We found that for longer filaments with wider lobe spacings, the acoustic guides were even less stable. Unless the 4-filament lobes were well balanced in energy and transverse position, the sound wave superposition would not form a well-defined air waveguide core. This is why a shorter 10 cm filament was used for the acoustic guide experiment. While the acoustic superposition guide is a promising approach, future experiments will need filaments generated by very well-balanced multi-lobe beam profiles.

By comparison, the thermal guides were far more robust, stable, and long-lived. Results from the thermal guide produced by a 70 cm long 4-filament are also shown in Fig. 4, where optimal coupling was found for *f/#* =200, in rough agreement with the earlier fibre-based estimate. An out of focus end mode image (not to scale) is shown to verify the presence of the thermal guide's lower density moat, whose lifetime here is much greater than in Fig. 3 owing to the initial bigger lobe spacing. Guided output modes as function of injection delay are shown imaged from a plane past the end of the guide, in order to minimize guide distortion of the imaging. These mode sizes are larger than upstream in the guide where 4-filament lobe spacing is tighter, but where we are unable to image reliably. We injected up to 110 mJ of 532 nm light, the maximum output of our laser, with 90% energy throughput in a single guided mode. This corresponds to a peak guiding efficiency (defined in the previous section) of 70%. Guiding efficiency vs. injected pulse delay is plotted in Fig. 4. As seen in that plot, peak guiding occurs at ~600 µs and persists out to ~2ms where the guiding efficiency drops to ~15%. Based on the guide core diameter of $2a$~150µm and the portion of the filament length with constant lobe spacing, $L$~50 cm, the guided beam propagates approximately $L\lambda/(\pi a^2)$ ~ 15 Rayleigh ranges.

**E. Simulations of waveguide development and guiding**

Owing to the linearity of the heat flow problem, the evolution of the 4-filament-induced density structure in the thermal regime can be calculated by finding the solution $T(x,y,t)$ to the 2D heat flow equation, $\partial T/\partial t = \alpha \nabla^2 T$, for a single filament source located at $(x, y)=(0, 0)$ and then forming $T_4(x,y,t) = \sum_{j=1}^{4} T(x-x_j, y-y_j, t)$, where $(x_j, y_j)$ are the thermal source locations in the 4-filament. Here $\alpha = \kappa/c_p$, where $\kappa$ and $c_p$ are the thermal conductivity and specific heat capacity of air. To excellent approximation, as shown in ref. [16], $T(x,y,t)$ is Gaussian in space. Invoking pressure balance, the 2D density evolution is then given by $\Delta N_4(r,t) = -N_b \frac{\Delta T_0}{T_b} \left( \frac{R_0^2}{R_0^2 + 4\alpha t} \right) \sum_{j=1}^{4} exp\left( \frac{-(x-x_j)^2 - (y-y_j)^2}{R_0^2 + 4\alpha t} \right)$, where $R_0$ is the initial $1/e$ radius of the temperature profile of a single filament and $\Delta T_0$ is its peak value above $T_b$, the background (room) temperature. Using $R_0$=50µm, $\Delta T_0$=15K, $\alpha$=0.21 cm$^2$/s for air [16], and source locations separated by 500µm, approximating our 70 cm 4-filament, gives the sequence of gas density plots shown in the upper panels of Fig. 5, clearly illustrating the development and persistence of the guiding structure over milliseconds.

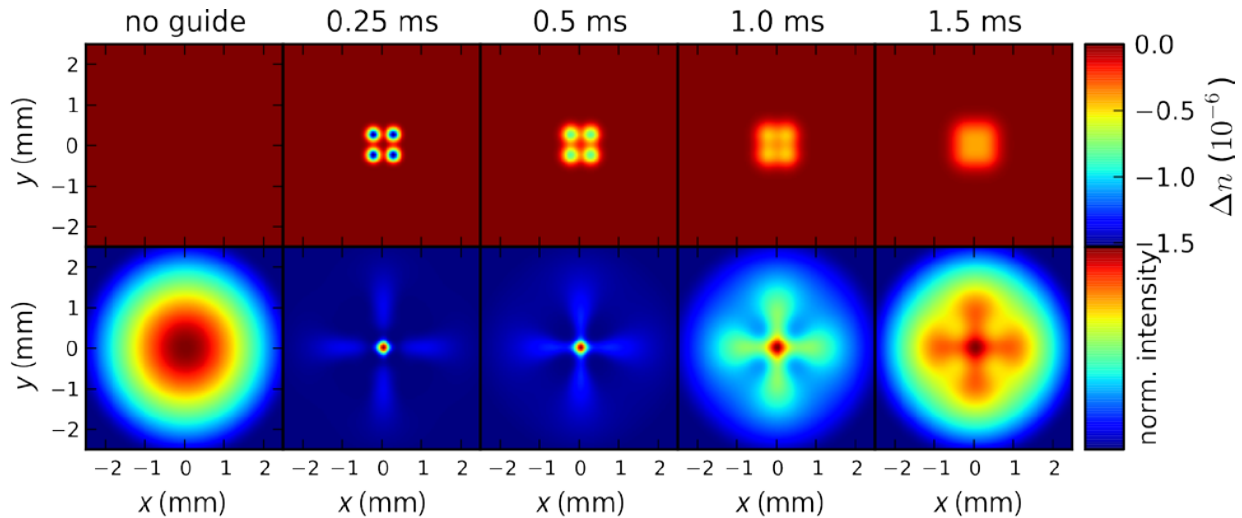

**Figure 5.** Simulation of the evolution of and guiding in a thermal air waveguide. The top row shows the index of refraction shift produced by the 4-filament-induced temperature profile as a function of time. The bottom row shows a BPM simulation of the guided laser beam profile at the end of a 70 cm waveguide produced by the 4-filament-induced refractive index change.

The propagation of the 532 nm beam in the waveguide was simulated in the paraxial approximation using the beam propagation method (BPM) [23]. The calculated intensity at the output of the waveguide is shown in the lower panels of Fig. 5. At early delays <100 μs, characteristics of a multimode waveguide are observed in the simulation, including mode beating. At later times, as the refractive index contrast decreases, the propagation is smoother, indicating single mode behaviour, consistent with the estimates using the fibre parameter. The simulation is in reasonable agreement with the experimental results. Axial nonuniformity in the waveguiding structure could explain the absence of four fold symmetry in the experimental data, whereas it is pronounced in the simulations.

### III. Conclusions

We have demonstrated the generation of very long-lived and robust optical waveguides in air, their extent limited only by the propagation distance of the initiating femtosecond filament array. This is ultimately determined by the femtosecond pulse energy used to ionize the gas. Based on a single filament diameter of ~100 μm, an electron density of ~$3\times10^{16}$ cm$^{-3}$ [4] and ionization energy of ~10 eV per electron, approximately 0.5 mJ is needed per metre of each filament. With a femtosecond laser system of a few hundred millijoules pulse energy, waveguides hundreds of meters long are possible.

What is the optical power carrying capacity of these guides? For ~10 ns pulses of the type used here for waveguide injection, the peak energy is limited by ionization threshold of $10^{13}$ W/cm$^2$ to ~20 J for our 150μm core diameter. However, the real utility of these air waveguides, in the thermal formation regime, derives from their extremely long millisecond-scale lifetime. This opens the possibility of guiding very high average powers that are well below the ionization threshold.

We now consider the robustness of our filament-induced waveguides to thermal blooming [12, 24] from molecular and aerosol absorption in the atmosphere. For thermal blooming, we consider the deposited laser energy which can raise the local gas temperature by a fraction $\eta$ of ambient, $P_g \Delta t / A = 1.5 \eta \alpha^{-1} p$, where $P_g$ is the guided laser power, $\Delta t$ is the pulse duration, $\alpha$ is the absorption coefficient, $A$ is the waveguide core cross sectional area, and $p$ is the ambient pressure. Thermal blooming competes with guiding when $\eta$ is approximately equal to the relative gas density difference between the core and cladding. In our measurements of the thermal air waveguide, the typical index (and density) difference between the core and cladding is of the order of ~2% at millisecond timescales. Taking $\eta$=0.02, $p$=1 atm, and $\alpha$=$2\times10^{-8}$ cm$^{-1}$ [12], gives $P_g \Delta t / A < \sim 1.5 \times 10^5$ J/cm$^2$ as the energy flux limit for thermal blooming. For example, for a 1.5 mm diameter air waveguide core formed from an azimuthal array of filaments, the limiting energy is $P_g \Delta t \sim 2.7$ kJ. Note that we use a conservative value for $\alpha$ at $\lambda$~1μm which includes both molecular and aerosol absorption for maritime environments [12], which contain significantly higher aerosol concentrations than dry air. If a high power laser is pulsed for $\Delta t$~2 ms, consistent with the lifetime of our 10 Hz-generated thermal waveguides, the peak average power can be 1.3 MW. It is possible that in such environments, air heating by the filament array itself could help dissipate the aerosols before the high power beam is injected, raising the thermal blooming threshold and also reducing aerosol scattering. An air waveguide even more robust against thermal blooming and capable of quasi-continuous operation may be possible using a kHz repetition rate filamenting laser. We have already shown that the cumulative effect of filamenting pulses arriving faster than the density hole can dissipate leads to steady state hole depths of order ~10% [16].

## Acknowledgements

The authors thank P. Sprangle and J. Palastro for useful discussions. This work is supported by the Air Force Office of Scientific Research, the National Science Foundation, the Department of Energy, and the Office of Naval Research.

## Appendix A: Simulation of filament-induced gas dynamics

The simulations of the gas hydrodynamic evolution is performed in cylindrical geometry using a one-dimensional Lagrangian one-fluid hydrocode, in which the conservation equations for mass, momentum and energy, $\partial \xi_i / \partial t + \nabla \cdot (\xi_i \mathbf{v} + \phi_i) = S_i$, were solved numerically. For the mass equation, $\xi_1 = \rho$ and $\phi_1$=0, for the momentum equation, $\xi_2 = \rho \mathbf{v}$ and $\phi_2 = P \vec{\mathbf{I}}$ (where $\vec{\mathbf{I}}$ is the unit tensor), and for the energy equation, $\xi_3 = \varepsilon + \frac{1}{2} \rho \mathbf{v}^2$ and $\phi_3$=$P\mathbf{v}$+$\mathbf{q}$. Here, $\xi$ is the volume density of the conserved quantity, $\phi$ is the flux of that quantity, and $S$ refers to sources or sinks, while $\rho$ is mass density, $\varepsilon$ is fluid internal energy density, $\mathbf{v}$ is fluid velocity, $P$ is gas pressure, and $\mathbf{q}= -\kappa \nabla T$ is the heat flux, where $\kappa$ and $T$ are the gas thermal conductivity and temperature. At all times, $S_1$=$S_2$=0 by mass and momentum conservation, but without approximations, $S_3 \neq 0$ because the thermal part of the energy density is changed by laser heating and by ionization/recombination of all the relevant species in the gas. However, we recognize that at times >~10 ns after laser filament excitation, all of the energy initially stored in free electron thermal energy and in the ionization and excitation distribution is repartitioned into a fully recombined gas in its ground electronic state. The 'initial' radial pressure distribution driving the gas hydrodynamics at times >10 ns is set by the initial plasma conditions $P_0(r) \approx (f_e / f_g) N_e(r) k_B T_e(r)$, where $k_B$ is Boltzmann's constant, $N_e(r)$ and $T_e(r)$ are the initial electron density and electron temperature profiles immediately after femtosecond filamentation in the gas, and $f_e$ and $f_g$ are the number of thermodynamic degrees of freedom of the free electrons and gas molecules. Here, $f_e = 3$, and $f_g = 5$ for air at the temperatures of this experiment ($\Delta T_0 < 100K$). To simulate the neutral gas response at long timescales we solve the fluid equations for the $\xi_i$, using $S_3=0$ and the initial pressure profile given by $P_0(r)$ above.

**Appendix B: Optical setup**

A $\lambda$=532 nm, 7 ns duration beam counter-propagates along a femtosecond filament structure generated by a 10 Hz Ti:Sapphire laser system producing $\lambda$=800 nm, 50 fs pulses up to 100 mJ. See Fig. 2(a). The optical arrangement is similar to our earlier experiment of ref. [16]. Here, the 532 nm pulse serves as either a low energy interferometric probe of the evolving gas density profile, using a folded wavefront interferometer, or as an injection source for optical guiding in the gas density structure. The 2D density profiles were extracted from the interferograms as described in ref. [16]. The delay of the 532nm probe/injection pulse is controlled with respect to the Ti:Sapphire pulse with a digital delay generator. The pulse timing jitter of <10 ns is negligible given the very long timescale gas evolution we focus on. For the injection experiments, up to 110 mJ is available at 532nm. The Rayleigh side-scattering image of Fig. 2(b) was obtained by concatenating multiple images from a low noise CCD camera translated on a rail parallel to the filament. The images were taken through an 800 nm interference filter.